\documentstyle[12pt]{article}
\newcommand{\beq}{\begin{eqnarray}}   
\newcommand{\eeq}{\end{eqnarray}}

\newcommand{\bq}{\begin{eqnarray}}   
\newcommand{\eq}{\end{eqnarray}}

\newcommand{\gsim}{\lower.7ex\hbox{
\;\stackrel{\textstyle>}{\sim}\;$}}
\newcommand{\lsim}{\lower.7ex\hbox{$
\;\stackrel{\textstyle<}{\sim}\;$}}

\begin{document}
\begin{titlepage}
\renewcommand{\thefootnote}{\fnsymbol{footnote}}

\begin{center} \Large
{\bf Theoretical Physics Institute}\\
{\bf University of Minnesota}
\end{center}
\begin{flushright}
OUTP-98-50P\\
TPI-MINN-98/05-T\\
UMN-TH-1708-98\\
hep-ph/9807286
\end{flushright}
\vspace*{1cm}

\begin{center}
{\Large \bf Chiral Symmetry Breaking without Bilinear Condensates, 
Unbroken Axial $Z_N$ Symmetry, and Exact QCD
Inequalities}

\vspace{0.8cm}

{\Large Ian I. Kogan,    Alex  Kovner}

\vspace{0.1cm}

{\it  Theoretical Physics, Oxford University. 1 Keble Road, Oxford 
OX13NP, 
UK}
\vspace{0.2cm}

 {\it and}

\vspace{0.5cm}

{\Large  M.  Shifman} 

\vspace{0.1cm}
{\it  Theoretical Physics Institute, Univ. of Minnesota,
Minneapolis, MN 55455, USA}

\end{center}

\vspace*{.3cm}

\begin{abstract}

An alternative pattern  of the chiral symmetry breaking, suggested 
recently, is investigated. It could be self-consistent provided  
that
the chiral SU$(N_f)\times$SU$(N_f)$ symmetry is broken spontaneously
down to  SU$(N_f)\times Z_{N_f}$  rather than to SU$(N_f)_V$. 
The discrete axial $Z_{N_f}$ then would play a custodial role preventing
the quark bilinears from condensation. It is shown that 
this pattern of the chiral
symmetry breaking is ruled out in QCD by  exact inequalities. 
It is not ruled out, however, in other gauge theories with 
scalar quarks and/or Yukawa couplings. 

\end{abstract}

\vspace{1cm}

\end{titlepage}

The lessons from supersymmetric gauge theories \cite{Nati} teach us 
that
gauge theories admit a broad spectrum of various dynamical regimes 
and, in 
particular, various patterns of the chiral symmetry breaking (CSB). 
Since QCD is not solved theoretically it is not unreasonable
to search for viable alternatives to the standard CSB pattern ascending 
to the 
famous work of Gell-Mann, Oaks and Renner \cite{GOR}. Recently an 
alternative regime has been suggested  by Stern and collaborators 
who also performed an 
extensive analysis of possible phenomenological consequences
\cite{stern1}. The basic elements of this scenario are as follows.
Consider QCD with $N_f$ massless flavors.
Then $N_f^2-1$ flavor-nonsinglet axial currents are coupled
to $N_f^2-1$ massless pions,
\beq
 \langle 0| A^a_\mu | \pi^b \rangle = i q_\mu \, F_\pi \, \delta^{ab}\, ,
\eeq
where 
$$
A^a_\mu = \bar\psi \frac{\lambda^a}{2} \, \gamma_\mu\gamma_5 
\psi\, , 
\,\,\,
$$
$\lambda^a$ are analogs of the Gell-Mann matrices acting in the flavor 
space 
(the color indices are suppressed), $q$ is the pion momentum, and 
$F_\pi$ is 
the coupling constant, $F_\pi\neq 0$. 
Correspondingly, the  two-point 
 current correlation
function
behaves in chiral limit   as prescribed by  the
Goldstone theorem
\begin{eqnarray}
i\int d^4x e^{iqx}\langle 0|A_\mu^a(x)\,\, A_\nu^b(0)|0\rangle  =
\left(
g_{\mu\nu} - {q_\mu q_\nu\over q^2}\right)
\delta^{ab}F(q^2); \ \ \ \ \ F(0) =  F_\pi^2\, .
\label{gold}
\end{eqnarray}
At the same time, unlike the standard scenario,  
the condensate 
\beq
S=\bar\psi\psi\equiv \sum_{f=1}^{N_f}\bar\psi_L^f\psi_R^f + {\rm H.c.}
\label{qbc}
\eeq
does not develop,
$\langle S\rangle = 0$. At $m_q\neq 0$ but $\ll \Lambda$ this would 
imply 
that  (see below)
\beq
\langle S\rangle = O(m_q )\, \,\,\, \mbox{and}\, \,\,\, m_\pi^2 = O(m_q^2 
)
\, \,\,\, \mbox{and}\, \,\,\,  \langle 0 | P^a|\pi^b\rangle = 
\delta^{ab}\times 
O(m_q )\, , 
\label{count}
\eeq
in sharp contradistinction with the generally accepted picture 
where\footnote{The analysis of Ref.
\cite{stern1} is  somewhat  more general since it assumes the
chiral counting of Eq. (\ref{count}) but does not necessarily require
that $\langle S\rangle$ vanishes in the chiral limit. 
The situation with a nonvanishing but an unnaturally small condensate 
at
$m_q=0$ is logically possible. It is  even harder to find a
natural explanation for this kind of scenario, however,  and we do not 
consider
it in this paper.}

\beq
\langle S\rangle = O(m_q^0 )\, \,\,\, \mbox{and}\, \,\,\, m_\pi^2= O(m_q 
)
\,\,\,  \langle 0| P^a|\pi^b\rangle = \delta^{ab}\times O(m_q^0 )\, .
\eeq
Here $P^a = \bar\psi \gamma_5{\lambda^a\over 2} \psi$. 

It was argued  \cite{stern2} that this unorthodox pattern of CSB 
($\langle 
S\rangle$
vanishes,  but the chiral symmetry is broken) is self-consistent and 
is possible from the point of view of the
properties of the density of states of the massless Dirac operator.
 As is well known, a nonvanishing  condensate $\langle S\rangle $
 exists only when the density of states  in the spectrum of the Dirac
operator  is non-zero at the origin \cite{bc}. At the same time, the 
condition 
for
 non-zero $F_{\pi}$ is weaker,  and 
{\em a priori}  it is not ruled out 
\cite{stern2} that
the density of states
vanishes  at the origin (so that $\langle S\rangle =0$), but the pion 
constant 
$F_\pi$  does not. This would be sufficient for the chiral symmetry to be 
spontaneously broken.

If so, the ample phenomenological consequences would be
strongly different from the standard picture. In particular,
the value of the chiral condensate
in   QCD with $m_q\neq 0$ must  be much smaller than usually assumed 
\cite{stern1}. It was 
argued \cite{stern1} that the
 low-energy mesonic data available so far can be successfully described  
if the
condensate $\langle S\rangle $ 
is smaller by an order of magnitude 
than generally accepted,  $-N_f\times (250\, \mbox{MeV})^3$, 
provided the nonstrange quark
masses
are larger by a factor 3 to 4 than currently believed. The 
strange quark mass is also significantly larger.

In this work we further investigate the pattern of CSB suggested in
\cite{stern1,stern2} with the purpose of answering two questions:

\vspace{0.2cm}

(i) Is this regime self-consistent and, if so, why the spontaneous CSB
might not lead to the emergence  of the natural order parameter
$\langle S\rangle \neq 0$?

(ii) Is this regime attainable in QCD?

\vspace{0.2cm}

The answer to the first question is positive. We identify a pattern of CSB 
where a custodial discrete subgroup $Z_N$ of the axial SU($N_f$) 
survives and 
protects the theory from the condensation of the quark bilinears 
(\ref{qbc}). 
It does not preclude, however, the formation of quartic condensates, 
which are 
singlets with respect to the vector SU($N_f$), singlets with respect to the 
custodial $(Z_N)_A$, but are not invariant with respect to the axial 
SU($N_f$).
These quartic condensates become the  order parameters
of the lowest dimension that are allowed.
Then we address the second question and show that in QCD {\em per se}
the pattern suggested in \cite{stern1,stern2} cannot be realized.
The exact inequalities \cite{nussinov} expressing the most general 
properties 
of the theory (basically, the fact that the quarks are the Dirac fields 
coupled to 
the gluons vectorially), do not tolerate the pion pole in the axial-axial 
correlation function without a similar pole in the pseudoscalar-
pseudoscalar
correlation function. Thus, in the quest for the alternative scenario of 
CSB, Eq. (\ref{count}),  one 
must 
leave QCD and turn to its generalizations (say, with the scalar quarks 
and/or 
Yukawa couplings).

\vspace{0.2cm}

Let us start from the first question. 
Why does the situation look paradoxical and counter-intuitive?
 If the symmetry
is broken, there seems to be no reason for the vacuum expectation value 
of the
fermion bilinear
to vanish. After all, $S$   is the lowest-dimension local operator
non-invariant under the axial SU($N_f$). Then,  how can  one 
understand a
regime in which  it vanishes, and the order parameter for CSB  has  
higher 
dimension? The examples of such higher-dimensional operators
are
$$
O_1 = \bar\psi\, \frac{\lambda^a}{2}\, \gamma_\mu (1-\gamma_5) \psi 
\cdot
\bar\psi\, \frac{\lambda^a}{2}\, \gamma_\mu (1+\gamma_5) \psi
$$
and
\beq
  O_2 = \bar\psi\, \frac{\lambda^a}{2}\,  (1-
\gamma_5) 
\psi \cdot
\bar\psi\, \frac{\lambda^a}{2}\, (1+ \gamma_5) \psi\, .
\label{4dimop}
\eeq
The point  is that the bilinear operator $S$ and the
quadrilinear operators (\ref{4dimop}) have {\it different}
transformation properties under the chiral 
SU$_L(N_f)\times$SU$_R(N_f)$.
In particular,
$\bar\psi_L\psi_R$ 
belongs to the fundamental representation
of 
SU$_R(N_f)$ and the antifundamental representation
of 
  SU$_L(N_f)$, whereas $O_{1,2}$ 
 transform as the  adjoint representation under both
SU$_L(N_f)$ and SU$_R(N_f)$. The expectation value of 
$\bar\psi_L\psi_R$ 
leaves intact the vectorial diagonal subgroup  of 
SU$_L(N_f)\times$SU$_R(N_f)$ and breaks completely the axial 
SU$(N_f)$.
At the same time, the operators $O_{1,2}$ are, additionally,  invariant 
under  the discrete $Z_N$  axial subgroup. In fact they are invariant
 under  the whole   $(Z_N)_L \times (Z_N)_R$ group
\begin{equation}
\psi_L\rightarrow e^{i{2\pi n\over N_f}}\psi_L,
 ~~~ \psi_R\rightarrow  e^{i{2\pi m\over N_f}}\psi_R
\label{zn1}
\end{equation}
For $n=m$
the resulting transformation is the same as the center of $SU_V(N)$,
which is unbroken in any case, and, therefore, it is sufficient to
consider either left or right  transformations  only. In the following
 we will concentrate on $(Z_N)_{L}$
\begin{equation}
\psi_L\rightarrow e^{i{2\pi n\over N_f}}\psi_L,
 ~~~ \psi_R\rightarrow \psi_R 
\label{zn}
\end{equation}

The regime considered in \cite{stern1,stern2}
 is natural if the pattern of CSB
is not the conventional SU$_L(N_f)\times$SU$_R(N_f)\rightarrow$
SU$_V(N_f)$
but, rather, the unorthodox  
\beq
\mbox{SU}_L(N_f)\times \mbox{SU}_R(N_f)\rightarrow
\mbox{SU}_V(N_f)\times (Z_{N_f})_A\,.
\label{upcsb}
\eeq
The vanishing of $\langle S\rangle $ in the chiral limit
is then insured by the unbroken axial $Z_N$.
Note that  the current
correlation function \footnote{$\Delta_1$ obviously vanishes if
the chiral symmetry is unbroken; dynamical CSB generates a non-
vanishing
$\Delta_1$.}
$$
\Delta_1 =i\int dx e^{iqx} \langle \bar\psi (x)\, \frac{\lambda^a}{2}\, 
\gamma_\mu 
(1-\gamma_5)\psi(x) \, , 
\bar\psi (0)\, \frac{\lambda^a}{2}\, \gamma_\nu (1+\gamma_5) \psi 
(0)\rangle
$$
which exhibits the very same pole behavior as in Eq. 
(\ref{gold}),  is invariant under $(Z_N)_A$ but not under the 
SU($N)_A$. If $\Delta_1$ does not vanish in  the chiral limit, 
there are no reasons to expect that the local operators with the 
appropriate
symmetry properties (e.g. $O_{1,2}$) will have zero expectation values. 
In fact, 
the existence of a local order parameter is necessary if the
low-energy physics is to be described by an effective chiral
Lagrangian, as is assumed in \cite{stern1}.

Let us now verify   that the unorthodox symmetry breaking pattern
(\ref{upcsb})
implies the 
same chiral counting as in Ref.  
\cite{stern1}.
Consider the chiral limit $m_q=0$.
The first observation is as follows: since all the broken symmetry 
generators 
$A_\mu^a$ 
are invariant under the axial $Z_N$ transformations, so are the
Goldstone bosons -- the pions. On the other hand, the pseudoscalar 
density
operator $P^a$ is not invariant \footnote{$P^a$ transforms as a sum of
two irreducible representations $(N,\bar N)+(\bar N,N)$ under
SU$_L(N)\times$SU$_R(N)$. None of these two
has zero ``$N$-ality".}
under $(Z_N)_A$.
This means that in the chiral limit the pion does decouple
from  the pseudoscalar density operator, 
\begin{equation}
\langle0|P^a|\pi\rangle =0\, .
\end{equation}
Incidentally, as a consequence of this decoupling, 
the lowest excitation in the pseudoscalar channel is a massive meson.
This massive particle is absolutely stable. It can not decay into the
massless pions, since it carries nontrivial $Z_N$ quantum numbers,
whereas any state consisting of arbitrary  number of pions is  $Z_N$ 
singlet.

Switching on the quark mass term we observe that it violates the 
$(Z_N)_A$
symmetry explicitly, and, therefore, away from the chiral limit
a nonvanishing bilinear condensate will be induced. Its value will
be proportional to the quark mass
$\langle S\rangle \propto m_q$.
Correspondingly, the pion will not remain massless. However, its mass
will be much smaller than in the standard scenario, unless one adjusts
the quark masses appropriately, making them larger. Indeed,
 from the  Gell-Mann-Oakes-Renner relation \cite{GOR}
\begin{equation}
F_\pi^2m_\pi^2=-m_q\langle S\rangle
\end{equation}
it follows
that the square of the pion mass is proportional to the square of the 
quark mass rather 
than to the first power, $
m_\pi^2\propto m^2_q$. 
Moreover,  at nonzero $m_q$ the pion does not decouple from the 
pseudoscalar
density any more. This coupling is small,  however,  and proportional to
 the quark mass.
This is precisely the scaling laws assumed in \cite{stern1}.

At the qualitative level, it is clear that this regime can be sustained
from the phenomenological point of view only if the quark masses are
 much larger
than the standard values. Also the ratios 
$(m_u+m_d)/m_s$ and $(m_d-m_u)/(m_d+m_u)$
have to be smaller, since, otherwise, the isospin and the Gell-Mann SU(3) 
breaking 
in the hadron spectrum would be much more pronounced. For example, 
the splitting between the pion masses
would be of the order of the pion mass itself. All these features again 
are 
explicit in the analysis of \cite{stern1}.

As a side remark we note that 
if one wants to be fully consistent within the framework of this
scenario, 
in developing the low-energy chiral theory 
one     must 
use a slightly different effective Lagrangian compared to what is 
routinely 
used now.
The effective Lagrangian must  be a local functional of the order 
parameter 
which
is in the adjoint-adjoint representation of 
SU$_L(N_f)\times$SU$_R(N_f)$.
Therefore, the effective Lagrangian must  be constructed from a real
$(N^2_f-1)\times (N^2_f-1)$ matrix $V^{ab}=R(U)$, rather than 
from the 
unitary $N\times N$ matrix $U^{\alpha\beta}$.   Here $R(U)$ is the
adjoint representation of the SU(N) group element $U$. 
If one insists on writing the effective Lagrangian 
in terms of the matrix $U$, then  this Lagrangian will have  a local $Z_N$ 
invariance.
This is, of course, just a  manifestation of the fact mentioned above --
 $U$ does not couple to the pion if the quark mass vanishes.
Nevertheless, the effective Lagrangian ${\cal L }[V]$ describes massless
mesons and satisfies the same low-energy theorems as in the standard 
scenario.

At finite quark masses $U$ does mix with $V$ and, if the mixing 
is large enough, one can
certainly  use ${\cal L }[U]$ in developing the low-energy theory. 
However, 
this description would be valid only at large enough $m_q$.

The fact that the order parameter is an adjoint rather than
 fundamental SU$(N)$ matrix has some bearing on the topological
properties of the vacuum manifold. The manifold spanned by the adjoint
representation matrix is SU$(N)/Z_N$ rather than
$SU(N)$. Nevertheless, the third homotopy groups $\pi_3$ of SU$(N)$ 
and 
SU$(N)/Z_N$ 
are
the same, and the effective Lagrangian ${\cal L }[V]$, therefore, admits 
solitonic
solutions much in the same way as ${\cal L }[U]$. These solitons are, of 
course,
naturally identifiable  with baryons. 
The fifth homotopy groups $\pi_5$ of SU$(N)$ and SU$(N)/Z_N$ are also 
the 
same; hence,
the theory based on ${\cal L }[V]$ admits the Wess-Zumino term, so that 
the 
baryons can be consistently 
quantized as fermions. 

Note that the discrete $Z_N$ is the {\em maximal} axial subgroup
that could survive without making the pattern of CSB self-contradictory. 
The extensive phenomenological analysis of the existing  meson data in 
Ref. 
\cite{stern1} seemingly does not immediately rule out  the unorthodox 
regime 
of CSB. Moreover, as we have just seen, generally speaking it  is 
theoretically 
self-consistent and even appealing.
If this alternative regime were sustainable in actual QCD this would be
an  exciting possibility.
We pass now to the analysis of the second question --
whether or not the pattern (\ref{upcsb}) and the ensuing scenario are 
attainable
in QCD. 

The standard argument against the unbroken axial symmetry in QCD
is the parity doubling that would be present in the spectrum if the
axial symmetry were unbroken (then no massless pion appear).
The parity transformation does not commute with the left- and 
right-handed currents. Under parity
\begin{equation}
(V+A)_\mu\rightarrow (V-A)_\mu\, .
\end{equation}
That is why the unbroken SU$(N)\times$ SU$(N)$, together with the 
unbroken parity,
requires  degeneracy in the spectrum. The most striking
consequence would be the degeneracy between $\pi$(135) and 
$a_0$(980), as
well as $\rho$(770) and $a_1$(1260). The splitting between the
would-be chiral partners in both cases
is too large to be explained by a finite
quark mass, so  one is forced to conclude that chiral symmetry is
broken.
On the purely theoretical side, the regime with the unbroken axial 
SU$(N)$, parity doubling, and no massless pion is ruled out by the 't 
Hooft matching of the AVV anomalous triangles \cite{thooft1}.

Would the same argument exclude the unbroken axial $Z_N$? The 
answer is
not so clear-cut. 
Since now the massless pion is present, the 't Hooft matching is 
automatic. It is still true that the parity transformation
$\psi_L\rightarrow\psi_R$ does not commute with the $Z_N$
transformation (\ref{zn}). Unbroken $Z_N$ therefore still requires a
degeneracy between the lowest excitation in the scalar and in the
pseudoscalar channels in the chiral limit.
However, as we have discussed above, the lowest
excitation in the pseudoscalar channel would be not the massless pion,
but, rather, a stable massive particle. As a result, one must compare
$a_0$(980) with the first pseudoscalar excitation, excluding the
pion, that is $\pi$(1300). The splitting between these particles is
much less spectacular, and could, perhaps, be explained by finite quark
masses especially if we recall that the $Z_N$ unbroken 
phenomenology \cite{stern1}
requires quarks to be fairly heavy.
The degeneracy between $\rho$ and $a_1$ now is not required at all by 
$Z_N$,
since both, the vector and pseudovector currents, are invariant under 
$Z_N$.

The parity doubling constraint is avoided in the baryon spectrum in a
similar fashion.
One can construct several operators to interpolate the left- and
right-handed baryon states
\begin{eqnarray}
&&B^1_L=(\psi_L\psi_L\psi_R)_L, \ \ B^2_L=(\psi_L\psi_R\psi_R)_L, \ \ 
B^3_L=(\psi_L\psi_L\psi_L)_L\, ,\nonumber\\
&&B^1_R=(\psi_R\psi_R\psi_L)_R, \ \ B^2_R=(\psi_R\psi_L\psi_L)_R, \ \ 
B^3_R=(\psi_R\psi_R\psi_R)_R\, .
\end{eqnarray}
Here we have omitted the color and flavor indices; 
the subscript $L$ ($R$) means that the Lorentz indices in the
appropriate combinations are contracted in such a way that the resulting
state of three quarks is left- (right-) handed. For explicit
construction of these operators in terms of the quark fields see 
\cite{ioffe}.

Consider the case of three massless flavors. Then under the $Z_3$
transformation  (\ref{zn})
\begin{eqnarray}
&&B^3_{L,R}\rightarrow B^3_{L,R}\, ,\nonumber\\
&&B^1_L\rightarrow e^{i{4\pi n\over 3}}B^1_L,\ \ \ 
B^2_R\rightarrow e^{i{4\pi n\over 3}}B^2_R\, ,\nonumber\\
&&B^1_R\rightarrow e^{i{2\pi n\over 3}}B^1_R,\ \ \ 
B^2_L\rightarrow e^{i{2\pi n\over 3}}B^2_L\, .
\end{eqnarray}
The allowed mass terms that do not break parity and $Z_3$ 
invariance are
\begin{equation}
\bar B^3_LB^3_R+ \mbox{H.c.}\, ,\ \ \  \bar B^1_LB^2_R+\mbox{H.c.}\,  , \ 
\ \
\bar B^2_LB^1_R+\mbox{H.c.}\,  .
\end{equation}
It follows that the two states $B^1$ and $B^2$ should be
degenerate while $B^3$ is not required to have a partner. The
situation is  analogous to that with the
pseudoscalar-scalar meson pair. The lowest-lying baryon could be in
the $B^3$ channel and nondegenerate, while the approximate
degeneracy must  be present only between the higher mass 
states \footnote{With two light 
flavors the situation is a little more delicate since
$B^3_L$ is not invariant under $Z_2$. One would then have to
consider an additional left-handed operator constructed from three
right-handed quarks. This extra state would be degenerate with
$B^3$.},  $N$(1440) and $N$(1535). In fact the very same pattern of
the parity doubling in the baryon spectrum was discussed  long 
ago by Dashen \cite{dashen}, as a possible manifestation of an unbroken
$Z_N$ chiral symmetry.

Thus, the  parity doubling argument is not strong enough
to invalidate the possibility of the unbroken chiral $Z_N$
on phenomenological grounds. Are there purely theoretical arguments
that could kill the pattern (\ref{upcsb}) and exclude all fermion
 bilinear condensates? The answer is yes. Unfortunately,  this realization 
of CSB 
is in direct contradiction with the
exact QCD inequalities \cite{wein} (see also \cite{nussinov}).
Consider the following inequality of the Weingarten type
  for the correlators of 
fermionic bilinears
\begin{equation}
|\langle P^a(x)P^b(y)\rangle | >  |
\langle A^{\mu a}(x) A^{\nu b}(y)\rangle | 
\label{ineq}
\end{equation}
 valid for arbitrary $x$, $y$ and the indices $a,b,\mu$ and $\nu$.
The correlator of the axial currents contains the 
contribution of the Goldstone
mesons. In the chiral limit its asymptotics at large distances is
\begin{equation}
\langle A^{\mu a}(x) A^{\nu b}(y)\rangle \rightarrow \mbox{const}\, 
F_\pi^2\delta^{ab}\, \frac{g^{\mu\nu}(x-y)^2 - 4(x-y)^\mu (x-y )^\nu}{ 
(x-y)^6}\, 
, \,\,\, {|x-y|\rightarrow\infty}\, .
\end{equation}
In the conventional CSB scenario the correlator 
of the pseudoscalar
densities also contains the pion pole and has the large distance 
asymptotics
\begin{equation}
\langle P^a(x)P^b(y)\rangle \rightarrow \delta^{ab}
\frac{A}{ (x-y)^2}\, , \,\,\, 
{|x-y|\rightarrow\infty}\, ,
\end{equation}
where $A$ is a constant of dimension $m^4$ and  of the order of a
typical hadronic scale $\Lambda^4$.
The inequality (\ref{ineq}) is obviously satisfied. However, in the
$(Z_N)_A$ invariant phase the pion decouples from the pseudoscalar
density. 
The residue $A$ vanishes and the asymptotic behavior  of the 
pseudoscalar
propagator is
\begin{equation}
\langle P^a(x)P^b(y)\rangle \rightarrow  B(x-y)^{-2}e^{-M|x-y|}\, , \,\,\, 
{|x-
y|\rightarrow\infty}\, ,
\end{equation}
where $M$ is the mass 
of the lowest 
massive particle in the pseudoscalar flavor-nonsinglet channel.
Clearly at distances $|x-y|\gg M^{-1}$ the correlator of the axial 
currents becomes
larger than the correlator of the pseudoscalar density, and the inequality 
(\ref{ineq}) is violated.

The argument above is somewhat oversimplified; there is a subtle point 
that 
must be discussed separately. Indeed, the proof of all exact inequalities 
\cite{wein,nussinov}
implies that a regularization is performed -- a small mass term is added 
to the
quark fields -- so that we do not have to deal with ill-defined 
 massless
Dirac
determinants. The quark mass term, no matter how small it is, couples 
the 
pion to the pseudoscalar density ($\langle 0|P^a|\pi\rangle \neq 0$), and 
also makes the axial vector 
two-point functions decay exponentially. Since both, the pion 
mass 
and the matrix element of the pseudoscalar
density,
are linear in the quark 
mass we have
\begin{eqnarray}
\langle A^{\mu a}(x) A^{\nu b}(y)\rangle & \rightarrow &
 F_\pi^2\delta^{ab}\bigg[\frac{g^{\mu\nu}z^2 - 4z^\mu z^\nu}{ 
z^6}
+cg^{\mu\nu}{m_\pi^2\over
z^2}\bigg]
e^{-m_\pi|z|}\nonumber\\
&& {|z| = |x-y|\gg M^{-1}}\, ,
\label{axi}
\end{eqnarray}
and
\begin{equation}
\langle  P^a(x)P^b(y)\rangle \rightarrow
{m_\pi^2C\over (x-y)^{2}}e^{-m_\pi|x-y|}\, , \,\,\,\, {|x-y|\gg M^{-1}}\, ,
\end{equation}
Here $c$ is a numerical constant of order $1$ and $C$ is a constant of
dimension $m^2$.

Now, at asymptotically large distances $|x-y|\to\infty $ the inequality 
(\ref{ineq})
could be satisfied,
depending on the exact numerical values of $c$ and $C$. 
A relation between $c$ and $C$ ensuring that the inequality is
satisfied at $|x-y|\to\infty $ is given in \cite{latorre},
where the limit $m_\pi z\gg 1$ was considered. Thus, by analyzing the 
exact inequalities in this limit one neither rules out nor confirms
the CSB pattern presented in Eq. (\ref{count}). 
Observe, 
however,
that the inequality 
(\ref{ineq})
must be valid at {\em any} values of $|x-y|$. 
Since  $m_\pi$ is arbitrarily small we can always find such values of 
$|x-y|$ 
in the range $M^{-1}\ll |x-y|\ll m_\pi^{-1}$ that the inequality is violated
 since for small enough $m_\pi$ there is a region where 
$m_\pi|x-y| \ll 1$ and the 
$(x-y)^{-4}$
term in the correlator of the axial currents  dominates \footnote{We
neglect the three-pion, five-pion, {\em etc.}  cuts  in Eq. (\ref{axi}) since 
they are suppressed by 
 powers of $1/M^2x^2$ and, therefore, irrelevant in the
region of distances we are interested in.}. One can  go even further
 and use these inequalities  to  obtain  a limit on a minimal possible 
value of the 
 bilinear condensate in actual QCD (with $m_q\neq 0$) 
since the constant $C$ is related to $\langle \bar{\psi}\psi\rangle $. It 
can not
 be too small -- otherwise the axial correlator will start, again, to 
dominate
 at intermediate distances. 

We, thus, conclude that the phase with the unbroken axial $Z_N$ 
symmetry
violates the exact QCD inequalities and, therefore, is ruled out as a 
vacuum phase 
of chirally invariant QCD
\footnote{An attempt to rule out the $(Z_N)_A$-symmetric phase
has been undertaken recently in \cite{knecht}. The authors analyze the 
Weinberg sum rules (i.e. the correlation function $\Delta_1$) in the large  
$N_c$ limit.
Integrals over the spectral density are expressed in  terms of a sum of 
local multi-quark operators. By applying a certain model of the 
resonance saturation it is possible to show  \cite{knecht} that at least 
some of these condensates must be nonvanishing. Needless to say that 
all local operators that enter the game must be $(Z_N)_A$-invariant, 
because so is 
the
correlation function $\Delta_1$.  An additional input is
provided by the large $N_c$ limit which implies 
factorization of the multi-quark operators and the emergence of
some bilinear condensates of the type
$\langle \bar{\psi} GG...G \psi\rangle$. For instance, the leading 
dimension-six operator in $\Delta_1$ is
\cite{SVZ} $\alpha_s (\bar u_L \gamma_\mu t^a u_L -\bar d_L 
\gamma_\mu t^a d_L )^2 \sim \alpha_s [(\bar u_L  u_R \bar u_R  u_L)
+ (u\to d)]$.
It is explicitely $(Z_N)_A$- invariant. If its vacuum expectation value 
does not vanish, and the naive $N_c$ counting rules take place,
one has to conclude that $\langle \bar u_L  u_R \rangle \neq 0$,
and $(Z_N)_A$ is broken \cite{knecht}. Thus the $(Z_N)_A$-symmetric 
scenario, with the quartic order parameters (\ref{4dimop}) and no 
bilineras, 
makes sense only at finite $N_c$, or if the naive $N_c$ counting rules are 
violated. Our consideration does not depend on the value of $N_c$.
}
. In 
fact the last argument suggests that even at finite 
quark masses the situation with small bilinear quark condensate is 
precarious.
As was mentioned above, the residue of the pseudoscalar density and 
hence the
condensate cannot be too small in order the QCD inequalities  to be
satisfied
at all distances. Although one can not rigorously  rule out 
the scenario suggested 
in \cite{stern1} on these grounds at finite quark masses, it seems to us 
that 
these considerations make the possibility of realization of this scenario 
in 
QCD
extremely improbable.

As a final remark, we note that the same argument -- exact  QCD 
inequalities 
--
is 
applicable
at finite temperatures since neither of the assumptions used in their 
derivation
requires $T=0$. Therefore, the situation with incompletely broken axial
symmetry is forbidden  at finite temperatures too. However, the exact 
QCD 
inequalities \cite{nussinov} are
not provable at finite {\em chemical potential}. It is  conceivable that
at finite density the phase with incompletely broken axial symmetry 
can be realized.
Recently  some interesting analysis of  QCD at finite 
density has been carried out \cite{krishna}.
So far, the calculations have been only performed in the mean field 
approximation
taking into account only the possibility of the  bilinear condensates. 
Nevertheless, in the
case of three flavors, the condensate favored by these calculations 
at high density
does not break axial symmetry completely. We hasten to add, though,  
that 
the 
symmetry breaking
pattern discussed there is 
certainly different
from the one we dealt with in the present paper. The unbroken
discreet group in \cite{krishna} is a subgroup of the axial $U(1)$. 
It would be 
interesting to explore
whether the phase of the type discussed here can be stable at 
finite density.

\vspace{0.3cm}

{\bf Acknowledgments}: \hspace{0.2cm} 
We are indebted to J. Wheater and C. Korthals-Altes 
for interesting discussions. 
We are grateful to M. Knecht, K. Rajagopal and J. Stern 
 for very informative and useful
correspondence and their comments on the manuscript.
This work was done in part when two of the 
authors (I.K. and A.K.) were visiting   Theoretical Physics Institute, 
University 
of Minnesota. They are grateful for kind hospitality.

This work was supported in part by DOE under the grant number
DE-FG02-94ER40823.
A.K. is supported by PPARC advanced fellowship.


\vspace{0.2cm}

\begin{thebibliography} {99}

\bibitem{Nati}
N. Seiberg, {\it Phys. Rev.} {\bf D49} (1994) 6857; {\it Nucl. Phys.}
{\bf B435} (1995) 129. 

\bibitem{GOR} M. Gell-Mann, R.J. Oakes, and B. Renner,
 {\it  Phys. Rev.} {\bf 175} (1968)  2195.

\bibitem{stern1} M. Knecht and J. Stern,
in {\em The Second  DAPHNE Physics Handbook},  Eds. L. Maiani, G. 
Pancheri and N. Paver (Frascati, 1995) p. 169
[hep-ph/9411253];
J. Stern, hep-ph/9712438. 

\bibitem{stern2} 
J. Stern, hep-ph/9801282.
 
\bibitem{bc} 
T. Banks and A. Casher, {\it Nucl. Phys.} {\bf B 168} (1980)
 103.

\bibitem{thooft1}
 G. 't Hooft,  in  {\it Recent Developments in Gauge Theories},
 Eds. G. 't Hooft {\em et al.},  (Plenum Press, New York, 1980).

\bibitem{ioffe}
B.L. Ioffe,
{\it Nucl. Phys.} {\bf B188} (1981) 317; {\it Z. Phys.} {\bf C18} (1983) 
67. 

\bibitem{dashen} 
R. Dashen, {\it Phys. Rev.} {\bf D183} (1969) 1245;

\bibitem{wein}
 D. Weingarten, {\it Phys. Rev. Lett.} {\bf 51} (1983)
1830.

\bibitem{nussinov}
E. Witten,  {\it Phys. Rev. Lett.} {\bf 51} (1983) 2351; 
S. Nussinov, {\it Phys. Rev. Lett.} {\bf 52}   (1984) 966;
D. Espriu, M. Gross and J.F. Wheater,  {\it Phys.  Lett.} {\bf 146 B} (1984)
 67.

\bibitem{latorre}
J. Comellas, J.I. Latorre and J. Taron {\it Phys. Lett.}
 {\bf B360} (1995) 109.

\bibitem{knecht} 
M. Knecht and E. de Rafael, {\it Phys. Lett.} {\bf
B424} (1998) 335.

\bibitem{SVZ}
M. Shifman, A. Vainshtein and V. Zakharov,
{\it Nucl. Phys.} {\bf B147} (1979) 385 (see Eq. (4.26)).

\bibitem{krishna} 
M. Alford, K. Rajagopal and F. Wilczek, hep-ph/9711395, 
~hep-ph/9804403; R. Rapp, T. Sch\"{a}fer,  E.  Shuryak and 
 M. Velkovsky, hep-ph/9711396; G.W. Carter and D. Diakonov, 
 hep-ph/9807219.


\end{thebibliography}
\end{document}